\documentclass[twocolumn,superscriptaddress,floatfix,preprintnumbers, nofootinbib,hyperref]{revtex4-2} 
\pdfoutput=1
\usepackage[colorlinks=true,breaklinks=true]{hyperref}
\usepackage[normalem]{ulem}
\usepackage[utf8]{inputenc}
\hypersetup{allcolors=[rgb]{0.0 0.0 0.6},linkcolor=[rgb]{0.75 0.05 0.05}}
\usepackage{amsmath,amssymb}
\usepackage{epsfig}  
\usepackage{graphicx}   
\usepackage{slashed}       
\usepackage{tikz}
\usepackage{tikz-feynman}
\usepackage{url}
\usepackage{color}
\usepackage{multirow}
\usepackage{comment}
\usepackage{amssymb}
\usepackage{orcidlink}
\usepackage[version=3]{mhchem}
\DeclareUnicodeCharacter{2212}{-}

\DeclareMathOperator{\cm}{cm}
\DeclareMathOperator{\GeV}{GeV}
\DeclareMathOperator{\eV}{eV}
\DeclareMathOperator{\meV}{meV}

\DeclareMathOperator{\MeV}{MeV}

\DeclareMathOperator{\s}{s}

\DeclareMathOperator{\erg}{erg}

\DeclareMathOperator{\km}{km}

\DeclareMathOperator{\kpc}{kpc}
\DeclareMathOperator{\g}{g}

\allowdisplaybreaks

\bibpunct{[}{]}{,}{n}{}{,}
\definecolor{ForestGreen}{RGB}{34,139,34}

\begin{document}

\title{Getting the most on supernova axions}

\author{Alessandro~Lella~\orcidlink{0000-0002-3266-3154}}
\email{alessandro.lella@ba.infn.it}
\affiliation{Dipartimento Interateneo di Fisica  ``Michelangelo Merlin,'' Via Amendola 173, 70126 Bari, Italy}
\affiliation{Istituto Nazionale di Fisica Nucleare - Sezione di Bari, Via Orabona 4, 70126 Bari, Italy}%

\author{Pierluca~Carenza~\orcidlink{0000-0002-8410-0345}}\email{pierluca.carenza@fysik.su.se}
\affiliation{The Oskar Klein Centre, Department of Physics, Stockholm University, Stockholm 106 91, Sweden
}

\author{Giampaolo~Co'~\orcidlink{0000-0002-9613-5211}}\email{Giampaolo.Co@le.infn.it}
\affiliation{Dipartimento di Matematica e Fisica ``Ennio De Giorgi'', Universit\`a del Salento, Via Arnesano, 73100 Lecce, Italy}
\affiliation{Istituto Nazionale di Fisica Nucleare - Sezione di Lecce,
Via Arnesano, 73100 Lecce, Italy} 

\author{Giuseppe~Lucente~\orcidlink{0000-0003-1530-4851}}
\email{giuseppe.lucente@ba.infn.it}
\affiliation{Dipartimento Interateneo di Fisica  ``Michelangelo Merlin,'' Via Amendola 173, 70126 Bari, Italy}
\affiliation{Istituto Nazionale di Fisica Nucleare - Sezione di Bari, Via Orabona 4, 70126 Bari, Italy}%

\author{Maurizio~Giannotti~\orcidlink{0000-0001-9823-6262}}
\email{mgiannotti@barry.edu}
\affiliation{Department of Chemistry and Physics, Barry University, 11300 NE 2nd Ave., Miami Shores, FL 33161, USA}%

\author{Alessandro~Mirizzi~\orcidlink{0000-0002-5382-3786}}
\email{alessandro.mirizzi@ba.infn.it}
\affiliation{Dipartimento Interateneo di Fisica  ``Michelangelo Merlin,'' Via Amendola 173, 70126 Bari, Italy}
\affiliation{Istituto Nazionale di Fisica Nucleare - Sezione di Bari, Via Orabona 4, 70126 Bari, Italy}%

 \author{Thomas~Rauscher~\orcidlink{0000-0002-1266-0642}}
 \email{Thomas.Rauscher@unibas.ch}
 \affiliation{Department of Physics, University of Basel, Klingelbergstr. 82, CH-4056 Basel, Switzerland}
 \affiliation{Centre for Astrophysics Research, University of Hertfordshire, Hatfield AL10 9AB, United Kingdom}

\date{\today}
\smallskip

\begin{abstract}
Axion-like particles (ALPs) coupled to nucleons might be copiously emitted from a supernova (SN) core.
We extend existing  bounds on free-streaming ALPs to the case in which these are so  strongly-interacting with the nuclear matter to be trapped in the SN core.
For strongly-interacting ALPs, 
we also extend the bound
from the absence of an ALP-induced signal in Kamiokande-II neutrino detector at the time of SN 1987A.
We find that combining the different arguments, SNe  exclude values of ALP-nucleon coupling $g_{aN}\gtrsim10^{-9}$ for ALP masses $m_a\lesssim 1\,\MeV$. 
Remarkably, in the case of canonical QCD axion models, the SN bounds exclude all values of $m_a \gtrsim 10^{-2} \eV$. This result prevents the possibility for current and  future cosmological surveys to detect any signatures due to hot dark matter QCD axion mass.

\end{abstract}

\maketitle

\section{Introduction}
In recent years, there has been an intense activity in studying the axion-like particles (ALPs) interaction with hadrons.
The ALP-nucleon coupling, $g_{aN}$ {($N=n,\,p$ for neutrons and protons, respectively)}, is accessible through several experimental methods which are currently being investigated.
These include observations of specific lines in the solar ALP flux~\cite{CAST:2009jdc,Borexino:2012guz,Bhusal:2020bvx,DiLuzio:2021qct,Lucente:2022esm}, 
dedicated nuclear magnetic resonance experiments~\cite{JacksonKimball:2017elr}, and long-range force experiments~\cite{Arvanitaki:2014dfa}.
Furthermore, the axion-pion coupling is largely responsible for the thermal production of axions in the early Universe~\cite{Chang:1993gm,Hannestad:2005df,Archidiacono:2013cha} and may be accessible to the next generation of cosmological probes.   
This has motivated a large number of dedicated studies (see, e.g., \cite{Giare:2020vzo,DiLuzio:2021vjd,DEramo:2021psx,DEramo:2022nvb}).    
Arguably, the most efficient ALP laboratories for the ALP-hadron interactions are supernovae (SNe). 
Since the observation of the neutrino signal from SN 1987A~\cite{Kamiokande-II:1987idp,Hirata:1988ad,Bionta:1987qt,IMB:1988suc}, the SN core has been identified as a 
powerful source of ALPs coupled with nucleons, up to masses 
${\mathcal O} (100)$~MeV~\cite{Turner:1987by,Raffelt:1987yt,Mayle:1987as,Burrows:1988ah,Burrows:1990pk,Raffelt:1990yz,Engel:1990zd,Turner:1991ax,Hanhart:2000ae,Raffelt:1993ix,Keil:1996ju,Fischer:2016cyd,Carenza:2019pxu,Carenza:2020cis,
Fischer:2021jfm,Lella:2022uwi,Carenza:2023lci}. 

The characterization of the ALP emission from the hot and dense nuclear medium typical of a young SN
has revealed to be more complex than naively thought. 
In particular, it has been recently  realized that for typical conditions of a SN core the pionic Compton processes
${\pi+N \to N+a}$~\cite{Carenza:2020cis,Fischer:2021jfm} dominate the ALP production with respect to the 
nucleon-nucleon ($NN$) bremsstrahlung ${N+N\to N+N+a}$, accounted for in the original 
calculations~\cite{Carena:1988kr,Brinkmann:1988vi,Raffelt:1993ix,Raffelt:1996wa,Carenza:2019pxu}. 

\begin{figure} [t!]
\centering
    \includegraphics[width=1\columnwidth]{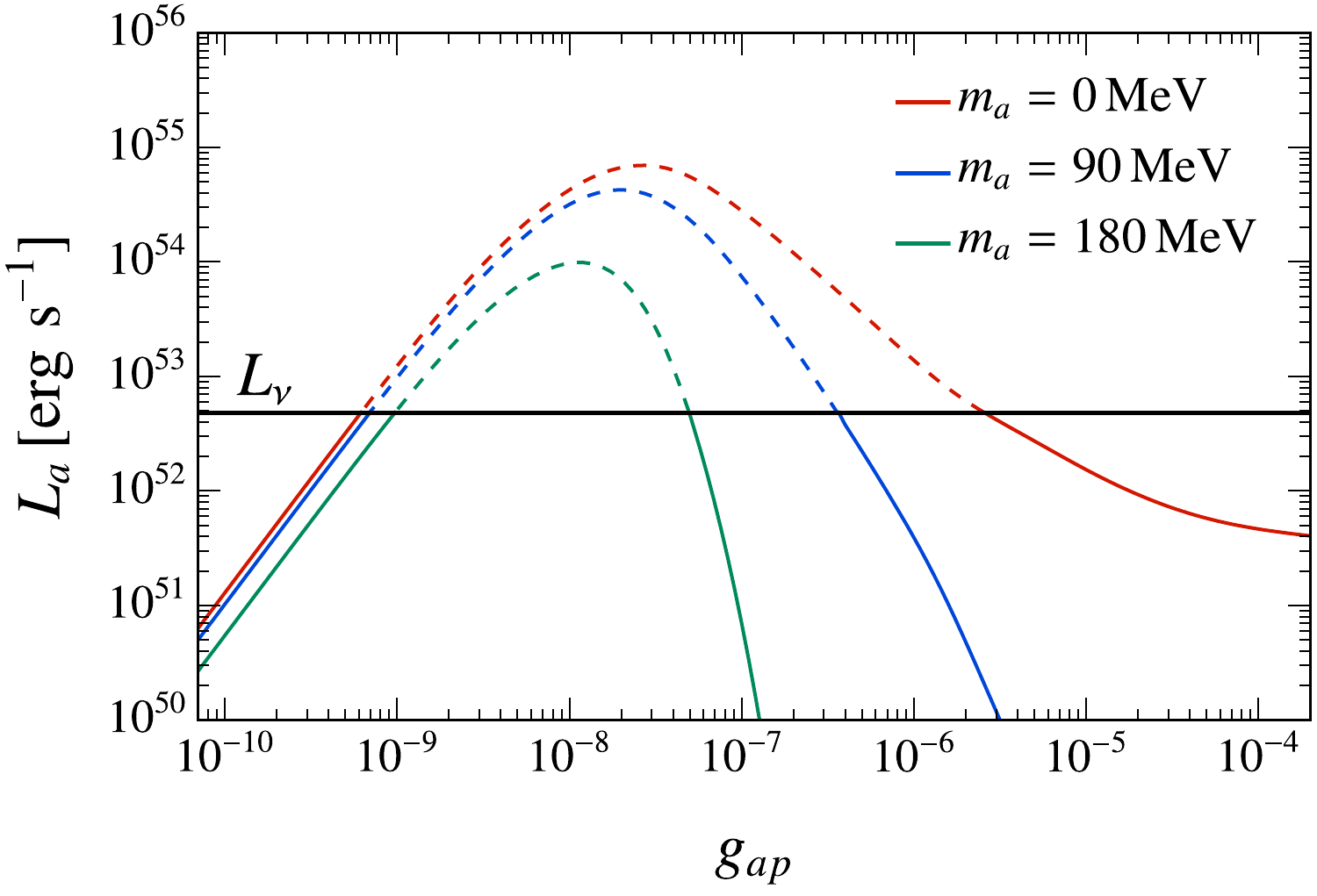}
    \caption{ALP luminosity at $t_{\rm pb}=1\,\s$ as a function of the ALP-proton coupling $g_{ap}$ for three different masses, assuming ${g_{an}=0}$. The black line sets the value of the neutrino luminosity. Dashed lines depict the range of couplings where the ALP luminosity exceeds the neutrino one.}
	\label{fig:lavsgap}
\end{figure}

ALPs weakly-coupled to matter, $g_{aN} \lesssim 10^{-8}$, are in the \emph{free-streaming regime} in which, once produced, they leave the star unimpeded, contributing to energy-loss and to the shortening of the observable SN neutrino burst~\cite{Raffelt:1987yt}. 
The most recent study shows that consistency with observations requires $g_{aN} \lesssim 8\times10^{-10}$ for 
$m_a \lesssim 10\,\MeV$~\cite{Lella:2022uwi}.
However, this argument cannot be extended to arbitrary large values of $g_{aN}$. 
For sufficiently large couplings, $g_{aN} \gtrsim 10^{-8}$, the SN environment becomes ``optically thick'' for the produced ALPs so that they cannot free-stream out of the star anymore and become trapped in the SN, analogously to neutrinos~\cite{Raffelt:1987yt}.  
This is known as the \emph{trapping regime}, studied in a number of seminal papers after the SN 1987A event~\cite{Raffelt:1987yt,Burrows:1990pk}. More recently, ALPs in this regime were considered in Ref.~\cite{Carenza:2019pxu}, which treated the free-streaming and trapping limiting cases separately. 
Despite its relevance, a study connecting properly these two regimes has never been performed. As we show below, a recently formulated ``modified luminosity criterion''~\cite{Chang:2018rso,Chang:2016ntp,Caputo:2021rux,Lucente:2020whw} allows the extension of the energy-loss argument from the free-streaming to the trapping limit, smoothly connecting the two regimes. Essentially, the modified luminosity criterion requires the ALP luminosity $L_a$ to be lower than the neutrino luminosity $L_\nu$ at post-bounce time $t_{\rm pb} =1~\s$, 
to avoid an excessive shortening of the neutrino burst. 
An example of how this method is applied  is shown in Fig.~\ref{fig:lavsgap}, 
where $L_a$ is plotted as a function of the ALP-proton coupling $g_{ap}$ for three values of the ALP mass and compared with $L_\nu$. These results have been obtained by assuming the ALP-neutron coupling $g_{an}=0$, in analogy with the Kim-Shifman-Vainshtein-Zakharov (KSVZ) axion model~\cite{Kim:1979if,Shifman:1979if} which we will adopt as our benchmark in the following.
Notably,  the ALP luminosity grows quadratically with $g_{ap}$ in the free-streaming regime. 
After reaching the luminosity peak,
ALPs enter the optically thick regime, in which their emission can be approximately described as a blackbody radiation from an \emph{axion-sphere}~\cite{Caputo:2022rca}, in close
analogy with the case of neutrinos. 
In this regime, for increasing values of $g_{ap}$  the radius of the emission surface $R_{a}$ increases, while the local temperature of the axion-sphere $T_a$ becomes smaller.
The combination of these effects leads the ALP luminosity $L_a {\sim R_a^2\,T_a^4}$ to approach a plateau at very high couplings. We highlight that this behaviour has been already observed in other works employing the modified luminosity criterion~\cite{Lucente:2022vuo}. On the other hand, ALPs with $m_a \gtrsim 10\,\MeV$ suffer an additional Boltzmann suppression which exponentially reduces the luminosity when $T_a$ decreases.

Finally, as originally pointed out in Ref.~\cite{Engel:1990zd} for sufficiently  large couplings, $g_{ap}\gtrsim10^{-7}$, the ALP burst might produce an observable signal in underground neutrino Cherenkov detectors.  Specifically,  ALPs would excite the oxygen nuclei in the detector, whose de-excitation would lead to the emission of observable photons. The non-observation of this signal in coincidence with SN 1987A allowed the exclusion of an additional range of ALP-nucleon couplings.

In this paper, we present an original and detailed study of the physics of trapped ALPs.  
In particular, in Sec.~\ref{sec:ALPproduction} we summarize the ALP production channels from nuclear matter in SNe and discuss how the predicted ALP spectrum changes from the free-steaming to the trapping regime. 
In Sec.~\ref{sec:ModifiedLum} we provide a smooth extension of the ALP luminosity to the trapping regime, 
with the goal of generalizing the SN cooling bound to the case of trapped ALPs. Our calculation  considers generic massive ALPs, thus broadening the analysis beyond the classic QCD ``hadronic axion'' case~\cite{Moroi:1998qs,Chang:1993gm}. In Sec.~\ref{sec:EventsKII} we obtain a bound on strongly-coupled ALPs from the absence of a signal associated with the ALP burst in Kamiokande-II (KII) detector at the time of SN 1987A. This last result is based on a new calculation of the ALP-oxygen interaction cross section, presented in a companion paper~\cite{Carenza:2023wsm}. 
Our results allow us to exclude large regions of the ALP parameter space, including the entire region accessible to the next generation of cosmological probes sensitive to axion mass. The robustness of these results is discussed in Sec.~\ref{sec:Uncertainties}, where we analyze the possible uncertainties associated with these constraints. 
In Sec.~\ref{sec:BoundQCDaxion} we translate constraints on the axion-nucleon coupling on the QCD axion mass for canonical QCD axion models.
Finally, in Sec.~\ref{sec:Conclusions} we summarize and conclude. An Appendix with details on the ALP emission calculation and the ALP-absorption mean free path follows.

\section{ALP production and absorption}
\label{sec:ALPproduction}
ALPs can be produced in a SN core by
$NN$ bremmstrahlung $N+N\to N+N+a$~\cite{Carena:1988kr,Brinkmann:1988vi,Raffelt:1993ix,Raffelt:1996wa,Carenza:2019pxu}  and by pionic Compton processes 
$\pi+N \to N+a$~\cite{Raffelt:1993ix,Keil:1996ju,Carenza:2020cis}. 
Recent $NN$ bremsstrahlung calculations introduce corrections to the naive one-pion-exchange approximation, including many-body effects on the nucleon dispersion relations in the medium and its finite lifetime due to multiple scattering~\cite{Carenza:2019pxu}. Pionic processes have also been re-evaluated recently in Ref.~\cite{Lella:2022uwi} and now include  also contributions from the contact interaction term~\cite{Choi:2021ign}, and from vertices associated to ALP-$\Delta$ coupling~\cite{Ho:2022oaw}. 
We refer to Ref.~\cite{Lella:2022uwi}  for the state-of-the-art calculation of the emissivities for these processes (see also Ref.~\cite{Carenza:2023lci} for a comprehensive review of the ALP production mechanisms). 

For sufficiently high values of $g_{aN}$ couplings, ALPs could be reabsorbed in the nuclear medium inside the dense  SN core, by means of reverse processes \mbox{$ N+N+a \rightarrow N+N$} and $  N+a\rightarrow N+\pi$. Starting from the expressions for the ALP production spectrum $d^2n_a/dE_adt$, introduced in Ref.~\cite{Lella:2022uwi}, one can derive the mean free path (MFP) $\lambda_a$ associated to these two processes, as illustrated in Appendix~\ref{app:trapping}.

The integrated ALP spectrum over a spherically symmetric SN profile can be calculated as~\cite{Chang:2018rso,Chang:2016ntp,Caputo:2021rux,Caputo:2022rca}  
\begin{equation}
    \frac{d^2N_a}{dE_a\,dt}=\int_0^{\infty}4\pi r^2 dr\left\langle e^{-\tau(E_a^*, r)}\right\rangle\,\frac{d^2 n_a}{dE_a\,dt}\,, 
\label{eq:spectrum}
\end{equation}
where $r$ is the radial position with respect to the center of the SN core and $\tau (E_a^*, r)$ is the optical depth at a given ALP energy and position.
Notice that the exponential term $\left\langle \exp[{-\tau(E_a^*, r)}]\right\rangle$ encodes the absorption effects over the ALP emission during the SN cooling and it is obtained by averaging over the cosine of the emission angle $\mu$~\cite{Caputo:2021rux}
\begin{equation}
    \left\langle e^{-\tau(E_a^*, r)}\right\rangle=\frac{1}{2}\int_{-1}^{+1}d\mu\,e^{-\int_0^\infty ds\, \Gamma_a\left(E_a^*,\sqrt{r^2+s^2+2rs\mu}\right)}\,,
\label{eq:tau}
\end{equation}
where 
\begin{equation}
    \Gamma_a(E_a,r)=\lambda_a^{-1}(E_a,r)\left[1-e^{-\frac{E_a}{T(r)}}\right]
\end{equation}
is the reduced absorption rate defined as in Ref.~\cite{Caputo:2021rux} and $T(r)$ is the temperature profile. The expression in Eq.~\eqref{eq:tau} is evaluated at  ${E_a^*=E_a\times\alpha(r)/\alpha(\sqrt{r^2+s^2+2rs\mu})}$, allowing us to take into account the gravitational redshift from the point of ALP production to the point of absorption, with $\alpha$ the lapse factor which embodies the gravitational effects.
This procedure allows us to take into account ALP emissions in any direction, including backward, at each location. We highlight that Eq.~\eqref{eq:spectrum} allows us to calculate the ALP spectrum in a unified framework which smoothly connects the free-streaming and trapping regimes. 
Such a recipe was missing in the previous literature.

\begin{figure} [t!]
\centering
    \includegraphics[width=1\columnwidth]{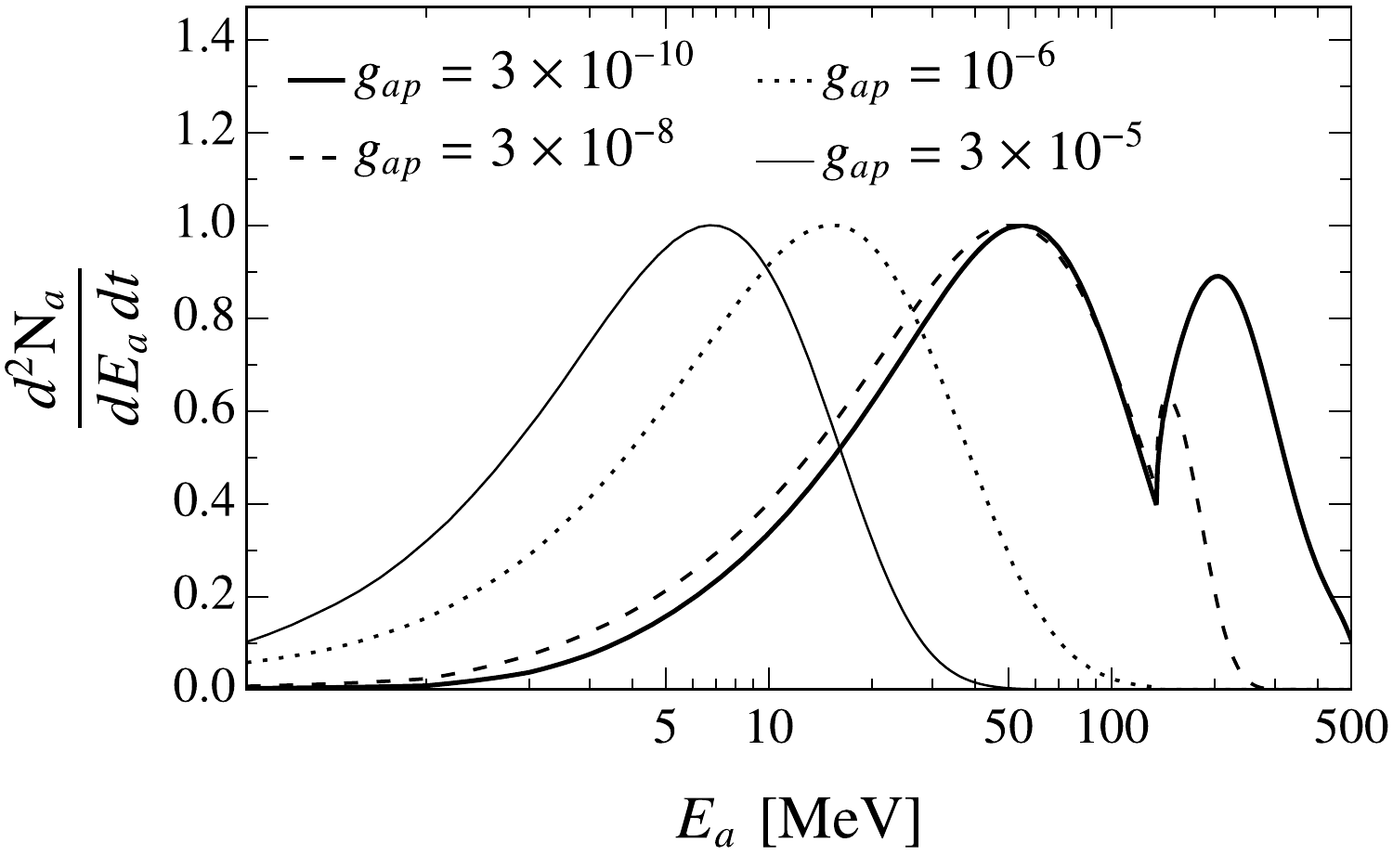}
    \caption{Normalized ALP spectrum in the massless case at $t_{\rm pb}=1\,\s$ for different values of the ALP-proton coupling $g_{ap}$ and for $g_{an}=0$.}
	\label{fig: Spectra}
\end{figure}

We have calculated these quantities numerically, using as reference the 1D spherical symmetric {\tt GARCHING} group's SN model SFHo-18.8 provided in~\cite{SNarchive}, launched from a stellar progenitor with mass $18.8~M_\odot$~\cite{2018ApJ...860...93S} and based on the neutrino-hydrodynamics code {\tt PROMETHEUS-VERTEX}~\cite{Rampp:2002bq}. The code takes into account all neutrino reactions identified as relevant for core-collapse SNe~\cite{Buras:2005rp,Janka:2012wk,2017PhRvL.119x2702B} and includes a 1D treatment of PNS convection via mixing-length theory convective fluxes~\cite{Mirizzi:2015eza} and muonic physics~\cite{2017PhRvL.119x2702B}. Our benchmark model, already used in previous analyses (see, e.g., Refs.~\cite{2020PhRvL.125e1104B,Caputo:2021rux,Caputo:2022mah,Fiorillo:2023frv}), leads to a neutron star (NS) with baryonic mass $1.351~M_\odot$ and gravitational mass $1.241~M_\odot$. We highlight that {\tt GARCHING} group's simulations do not account for the presence of pions in the SN core, since the pion properties in PNS matter are still under debate. Therefore, we estimated the pion chemical potential and a pion abundance by employing the procedure in Ref.~\cite{Fischer:2021jfm}, including the pion-nucleon interaction as described in Ref.~\cite{Fore:2019wib}. This post-processing addition of pions is justified as long as the impact of pionic matter on the PNS properties is not larger than the impact of muons. For our purpose, we employ a single snapshot of this SN model at $t_{\rm pb}=1\,\s$.\\
Fig.~\ref{fig: Spectra} displays the behaviour of the ALP spectrum in the massless case for different values of the ALP-proton coupling $g_{ap}$ at $t_{\rm pb}=1\,\s$. We  observe that in the free-streaming regime ($g_{ap}=3\times10^{-10}$, solid thick line) the spectrum clearly shows a bimodal shape with two peaks, one at ${E_a\simeq 50}$~MeV associated to $NN$ bremsstrahlung, and the other  one at $E_a \simeq 200$~MeV due to $\pi N$ process~\cite{Lella:2022uwi}. As the coupling grows both production and absorption processes become more efficient. 

At any given $g_{ap}$, inverse pion conversion (when kinematically allowed) is more efficient than bremsstrahlung, resulting in a dramatic reduction of the ALP MFP for $E_a\geq m_\pi$ (see Appendix~\ref{app:trapping}).
Then the part of the spectrum due to $\pi N$ process is suppressed because of pionic re-absorption that lowers the second peak of the spectrum for \mbox{$g_{ap}=3\times10^{-8}$} (dashed curve) till it 
washes it out completely for $g_{ap}=10^{-6}$ (dotted curve)  and $g_{ap}=3\times10^{-5}$ (solid thin curve). 

On the other hand, since the $NN$ bremsstrahlung is a thermal process, the first peak in the ALP spectrum reflects the  temperature of the regions where escaping ALPs are produced. 
In fact, as the ALP-nucleon coupling increases, ALPs become more and more trapped inside the inner regions of the SN core and the only ones able to escape are those produced in the outer layers, where the temperature is lower. 
Thus, the peak of the spectrum associated to $NN$ bremsstrahlung is shifted towards lower energies, from  $E_a\simeq 50\,\MeV$ 
for $g_{ap}=3\times10^{-10}$ in the free-streaming case down to $E_a\simeq 6\,\MeV$ for $g_{ap}=3\times10^{-5}$ in the strongly-coupled case, {where the emission takes place from a region at temperature $T\simeq4\,\MeV$.}

\section{Modified luminosity criterion}
\label{sec:ModifiedLum}
\begin{figure*}[t!]
\vspace{0.cm}
\includegraphics[scale=0.8]{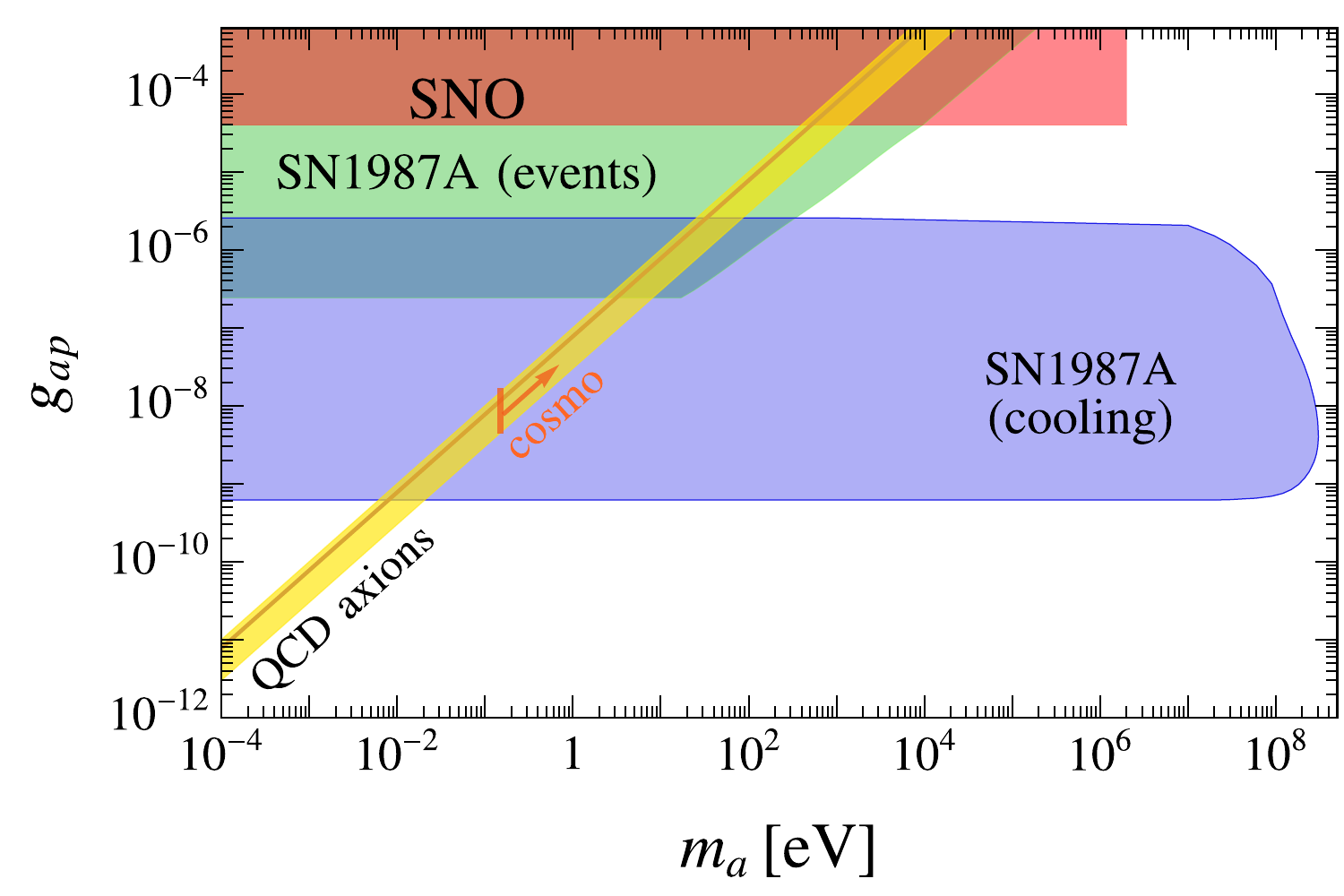}
\caption{Summary plot of the bounds in the $g_{ap}$ vs $m_a$ plane together with the QCD axion band (in yellow). 
The region labeled SNO is excluded by the search for $p+d\to \, ^3 {\rm He}+a\,(5.5\,{\rm MeV})$ solar ALP flux in SNO data~\cite{Bhusal:2020bvx}.  
The green and blue regions labeled SN1987A are ruled out from the non-observation of extra events inside the KII experiment and by the cooling argument. 
The orange line with the arrow within the QCD axion band shows the sensitivity of current and future cosmological experiments, $m_a\gtrsim0.15  \eV$~\cite{Archidiacono:2015mda,DEramo:2022nvb}. See the text for more details. }
\label{fig:CoolingBound}
\end{figure*}

The ALP luminosity can be computed as~\cite{Caputo:2021rux,Chang:2016ntp,Lucente:2020whw}

\begin{equation}
    L_a=\int_0^{R_\nu}4\pi r^2 dr\int_{m_a/\alpha}^\infty dE_a\, E_a \, \alpha(r)^2 \left\langle e^{-\tau(E_a, r)}\right\rangle\,\frac{d^2 n_a}{dE_a\,dt}\,, 
\label{eq:Luminosity}
\end{equation}
where the lower limit of integration $m_a/\alpha$ cuts away the fraction of heavy ALPs gravitationally trapped in the interior of the core~\cite{Caputo:2022mah,Lella:2022uwi,Lucente:2020whw}. Notice that in Eq.~(\ref{eq:Luminosity}) we have set upper integration limit of the radial coordinate at the neutrino-sphere radius $R_\nu$ in order to consider just the amount of energy carried away by axions which can impact the PNS cooling and the associated neutrino emission
duration. \newline
An excess in energy-loss during the SN cooling phase would have shortened the duration of the SN 1987A neutrino burst. Therefore, the 
modified luminosity criterion 
requires that at $t_{\rm pb}\sim1 \s$~\cite{Raffelt:1996wa,Chang:2016ntp} the ALP luminosity $L_a$ computed on the unperturbed model must not exceed the total neutrino luminosity $L_\nu$ provided by the same SN simulation. Namely, for our benchmark model, at $t_{\rm pb}=1\,\s$ we have
\begin{equation}
    L_a\lesssim L_\nu\simeq5\times10^{52}\,\erg \s^{-1}\,,
    \label{eq:CoolingCondition}
\end{equation}
This criterion allows us to exclude the blue region in  Fig.~\ref{fig:CoolingBound}. 
Note that in this case  we  exclude \mbox{$6\times10^{-10}\lesssim g_{ap}\lesssim 2.5\times10^{-6}$} for \mbox{$m_a\lesssim 10\MeV$} and \mbox{$7\times10^{-10}\lesssim g_{ap}\lesssim 3.5\times10^{-7}$} for \mbox{$m_a\sim\mathcal{O}(100)\MeV$}. 

The behaviour of the lower bound corresponding to free-streaming ALPs has already been discussed in Ref.~\cite{Lella:2022uwi}, so we will focus on the range $g_{ap} \gtrsim 10^{-8}$, where ALPs enter the trapping regime.
As discussed in the previous Section, strongly-coupled ALPs with $m_a<m_\pi$ can escape the SN core only if their energies are lower than the pion mass, where pionic production is not possible. Since the emission is mostly in the range $E_a\leq m_\pi$ the dominant absorption mechanism is inverse bremsstrahlung, which is less efficient than pionic absorption. Thus, high values of the ALP-proton coupling are required to saturate the condition in Eq.~(\ref{eq:CoolingCondition}) and the bound settles at $g_{ap}\simeq2.5\times10^{-6}$.
We observe that the bound updates the results in the trapping regime obtained in the previous literature~\cite{Chang:2018rso,Carenza:2019pxu}. 
In Ref.~\cite{Carenza:2019pxu} the ALP production and absorption took into account $NN$ bremsstrahlung which is the only relevant process in the trapping limit. The origin of the differences with the results obtained in this work are due to different motivations. Most importantly, the opacity is underestimated due to a miscalculation of the total MFP [see Eq.~(5.31) therein]. In addition, the bound in Ref.~\cite{Carenza:2019pxu} is obtained by assuming a different physics case ($g_{an}=g_{ap}$) and by employing a less refined criterion based on a rough calculation of the axion-sphere radius. On the other hand, in Ref.~\cite{Chang:2018rso} ALPs were constrained using the modified luminosity criterion, but employing  different emission and absorption rates, not accounting for all the corrections considered in this work and in Ref.~\cite{Lella:2022uwi}.
On the other hand, for $m_a\geq m_\pi$ pionic processes are possible for all energies. Consequently, the dominant absorption process in the entire range of energies is inverse pion conversion, and the ALP absorption rate is significantly enhanced reducing the ALP luminosity.
Then the constraint on $g_{ap}$ excludes a smaller region of the parameter space compared to the case of lower masses.
\newline

\section{ALP-induced events in Kamiokande-II detector}
\label{sec:EventsKII}
Strongly-coupled ALPs emitted during a SN explosion may lead to a detectable signal in large water Cherenkov neutrino detectors, as  proposed in the  seminal paper by Engel \emph{et al.}~\cite{Engel:1990zd}, in which the authors proposed to look for axion-induced excitation of oxygen nuclei with the subsequent emission of a photon to relax the system
\begin{equation}
    a+ \ce{^{16}O}\rightarrow \ce{^{16}O}^* \rightarrow \ce{^{16}O}+\gamma \,.
\end{equation}
A revised calculation of the cross-section for this process, using state-of-the-art nuclear models, is presented in Ref.~\cite{Carenza:2023wsm}.
However, on a dimensional ground we can estimate the order of magnitude of the oxygen absorption cross section as $\sigma\sim g_{ap}^{2}/m_{N}^{2}\simeq 10^{-46}\,{\rm cm}^{2}(g_{ap}/10^{-9})^{2}$, where $m_N = 938~\MeV$ is the nucleon mass. This estimate is in remarkably good agreement with the numerical results~\cite{Carenza:2023wsm}.\\
Then, for the calculation of the ALP-induced signal, we adopted a two-step approach similar to what was done in Refs.~\cite{Engel:1990zd,Langanke:1995he}. 
First, we calculated the population of ALP-induced excited states in $^{16}$O using the \emph{continuum random phase approximation} approach of Ref.~\cite{DeDonno:2016vge}. Then, in a second step, we evaluated the de-excitation spectrum of $^{16}$O$^*$ through $\gamma$-cascades and particle emission.
The emission of nucleons and $\alpha$-particles, as well as the $\gamma$-decays of their final nuclides, had also to be included because states above the particle separation energies in $^{16}$O$^*$ are initially populated. 
The $\gamma$- and particle-transitions were treated similarly to the calculation of the transmission coefficients in the statistical model reaction code SMARAGD~\cite{smargd}, as also described in Ref.~\cite{2011IJMPE..20.1071R}. 
Summed spectra comprising all emissions were obtained. We highlight that the numerical results for the cross section were derived by considering couplings inspired to the KSVZ axion model, which we adopted as our benchmark model. Finally, among the possible nucleon-nucleon interactions described in Ref.~\cite{Carenza:2023wsm}, in our estimates we employed the parametrization which gives the most conservative results.\\
Starting from the cross-section $\sigma$ and the ALP number flux $F_a$, it is possible to calculate the number of events in the detector following the recipe described in Ref.~\cite{Carenza:2023wsm}.
If ALPs were produced in the SN 1987A explosion, for sufficiently high values of $g_{ap}$, this flux would have been detected at the KII water Cherenkov detector~\cite{Kamiokande-II:1987idp,Hirata:1988ad}, as pointed out in Ref.~\cite{Engel:1990zd}. 
We remark that, our analysis, as well as the original one~\cite{Engel:1990zd}, focuses only on the KII detector, neglecting the SN neutrino detection at Irvine-Michigan-Brookhaven (IMB) observatory~\cite{Bionta:1987qt,IMB:1988suc}.
This choice is due to the different energy thresholds of the two detectors, with KII extending down to $E_{\mathrm{thr}}\simeq4\,\MeV$ sensitive to all the possible oxygen radiative de-excitation channels~\cite{Carenza:2023wsm}. 
On the other hand the IMB threshold $E_{\mathrm{thr}}\simeq19\,\MeV$ is too high to have a significant axion detection~\cite{Fiorillo:2023frv} in a region of the parameter space not already ruled out by other experimental constraints~\cite{Bhusal:2020bvx}. In particular, the energies of the photons from oxygen de-excitation lie mostly in the range $E_\gamma<15\MeV$ (see Ref.~\cite{Carenza:2023wsm} for further details), so that they result to be out of the reach of IMB detection.
In light of the most recent analysis of the SN~1987A neutrino burst, performed in Ref.~\cite{Fiorillo:2023frv}, we have assumed that the first nine events detected in KII during the first second are actually in good agreement with neutrino events expected from SN simulations.
Conversely, we do not include the latest three events at $t_{\rm pb} \sim 10$~s in our further considerations since  they are in tension with state-of-the-art simulations.\newline
During $2.7$ days around the SN 1987A time the background at KII was ${\overline{n}_{\rm bkg}\simeq0.02\,\,\text{events}/\s}$~\cite{Hirata:1988ad,Kamiokande-II:1987idp}. 
Therefore, one can exclude all values of $g_{ap}$ leading to
\mbox{$N_{\rm ev} \gtrsim 2\,\sqrt{\overline{n}_{\text{bkg}}\Delta t_a}$}
in the time window $\Delta t_a$ of the ALP signal. 
For ALPs with $m_a \lesssim 20$~eV~\cite{Raffelt:1996wa}, time-of-flight effects are negligible and one can assume that 
ALPs arrive in the same time window of neutrinos, so that  $\Delta t_a= \Delta t_\nu$. 
Using the results in Ref.~\cite{Carenza:2023wsm}, this leads to the constrain $g_{ap}\gtrsim 3\times 10^{-7}$.
For higher values of the ALP mass, ALPs would travel more slowly than neutrinos and their arrival on the Earth would be delayed with respect to the first neutrino event by $t=D\,m_a^2/2E_a^2$~\cite{Raffelt:1996wa},
where $D=51.4~\kpc$ is the SN 1987A distance to the Earth.
Furthermore, less energetic ALPs arrive later. 
Therefore, $\Delta t_a$ is shifted at later times with respect to $\Delta t_\nu$ and it is also spread due to the energy-dependent delay. Since the oxygen energy levels lie in the range
$E\in [9.55, 28] \,\MeV$, delays between the least and the most energetic ALPs detectable in KII can be estimated as

\begin{equation}
    \begin{split}
        \Delta t\,(m_a)&\approx t\,(E_{\text{min}},m_a)-t\,(E_{\text{max}},m_a) \\
        &\approx 1.82\s\,\left(\frac{m_a}{10\,\eV}\right)^2\,, \\
    \end{split}
\end{equation}
where $E_{\rm min} = 9.55~\MeV$, $E_{\rm max} = 28$~MeV and we have neglected the intrinsic dispersion of the ALP signal.
 Notice that for increasing ALP masses, the emission starts to be Boltzmann suppressed and the time delay rapidly increases.
From our analysis we have excluded the green region in Fig.~\ref{fig:CoolingBound}. 
For reference, we also show in red, the region excluded by the search for dissociation of deuterons induced by solar ALPs in the Sudbury Neutrino Observatory (SNO) data~\cite{Bhusal:2020bvx} ($g_{ap} \gtrsim4\times10^{-5}$ for $m_a\lesssim 1\,\MeV$).

\section{Uncertainties on SN bounds}
\label{sec:Uncertainties}
The SN cooling argument employed in this work to constrain the ALP parameter space may be affected by many uncertainties related to the SN modelling employed in the calculations. Indeed, the SN temperature and density profiles significantly affect the ALP emission rates, since they are very sensitive to the stellar properties. Therefore, this Section is devoted to discuss how robust our results are under variations of the SN inputs. In particular, we will focus on three important aspects: the presence or absence of pions inside the SN core; the dependence of the bounds on the used SN model and the possible ALP absorption by heavy nuclei in the outer SN envelope, potentially relevant for strongly-coupled ALPs.

\begin{figure*}[t!]
\centering
\makebox{
    \includegraphics[width=1.\columnwidth]{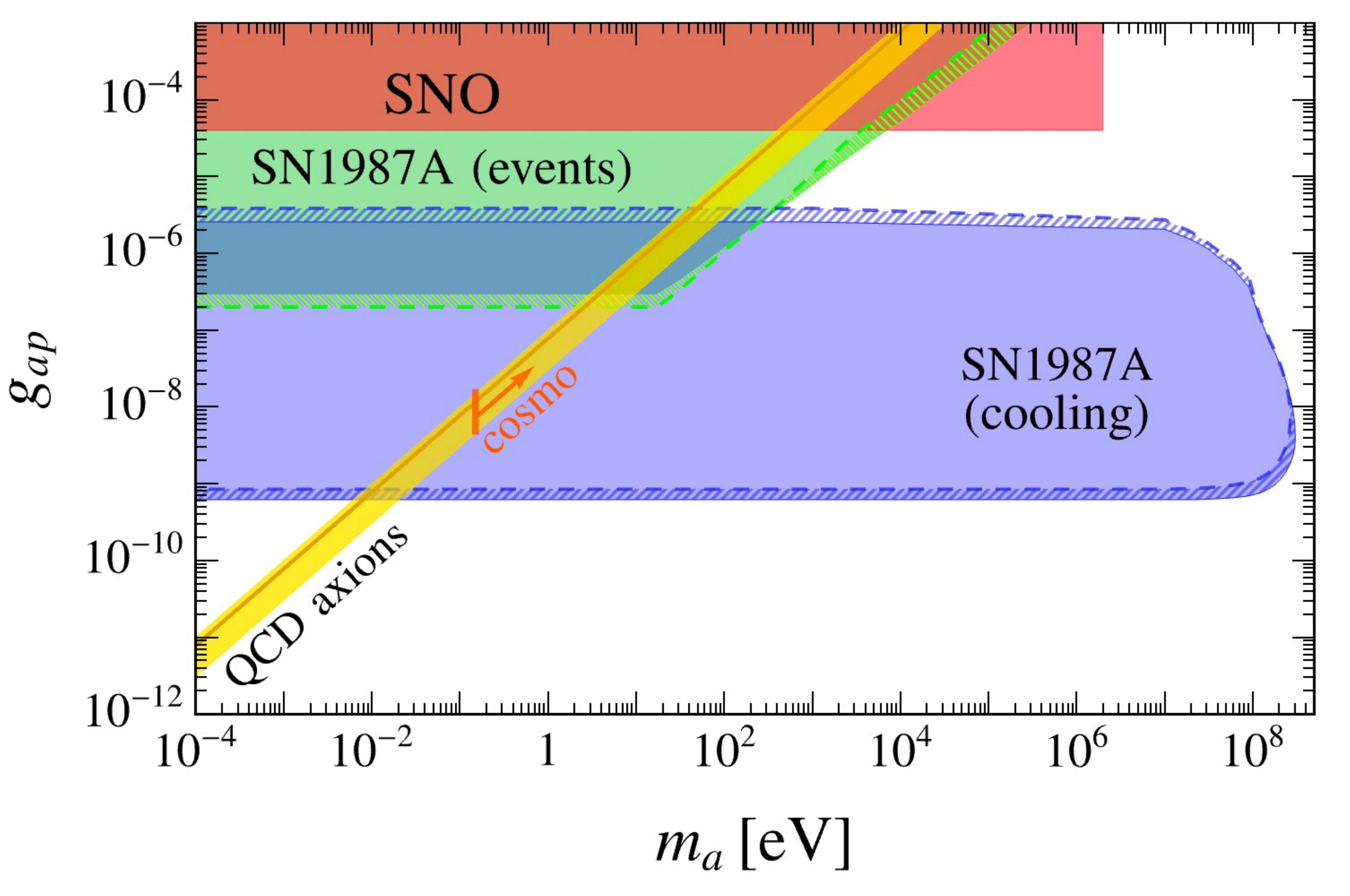}
       \hspace{0.3cm}
    \includegraphics[width=1.\columnwidth]{Figure4b.pdf}

    }
    \caption{Summary of the bounds with their uncertainties in the $g_{ap}$ vs $m_a$ plane together with the QCD axion band. 
    The color code is the same as Fig.~\ref{fig:CoolingBound}. {\it Left panel:} The hatched regions show the uncertainty on the bounds from the SN~1987A cooling (blue region) and the non-observation of extra events in the KII experiment in coincidence with SN 1987A (green region). In both the cases, solid thin lines delimit the region excluded by the {\tt GARCHING} model while the dashed thick lines delineate the one excluded using the {\tt AGILE-BOLTZTRAN} code. 
    {\it Right panel:}
    The blue hatched region showing the uncertainty on the cooling bound is obtained by assuming the presence of pions in the SN core, or not. The green hatched region displays uncertainties regarding ALP absorption by heavy nuclei in the neutrino-driven wind.
    }
    \label{fig:CoolingBound_withUncertainties}
\end{figure*}

\subsection{Dependence of the bounds on the SN models}
The SN emissivity of ALPs coupled to nuclear matter is very sensitive to the SN temperature and the density profiles, which can be significantly different from one SN model to another. Therefore, it is important to test the robustness of the constraints placed in this work under the change of the SN profile employed. As discussed in Sec.~\ref{sec:ALPproduction}, the results reported in the previous sections are obtained by using the {\tt GARCHING} SFHO-18.8 model. In the same way, we have also characterized ALP emission by employing a 1D spherically symmetric and general relativistic hydrodynamics SN model launched from a $18~M_{\odot}$ progenitor~\cite{Woosley:2002zz} and based on the {\tt AGILE-BOLTZTRAN} code~\cite{Mezzacappa:1993gn,Liebendoerfer:2002xn}. The latter code includes a full set of neutrino weak interaction processes~\cite{Fischer:2018kdt,Guo:2020tgx} but, contrary to the {\tt GARCHING} model, it does not account for convection effects. This SN model predicts a lower temperature and density in the SN core at $t_{\rm pb}=1~\s$ compared to the {\tt GARCHING} one, affecting the ALP emissivity computed in this work. \\
The left panel in Fig.~\ref{fig:CoolingBound_withUncertainties} shows the SN bounds obtained using the {\tt GARCHING} model (green and blue regions delimited by solid lines) and the {\tt AGILE-BOLTZTRAN} model (dashed lines). In particular, the lower part of the cooling bound is slightly relaxed by switching to the {\tt AGILE-BOLTZTRAN} code. This is essentially due to the fact that the {\tt GARCHING} simulation predicts higher temperatures in the inner regions of the SN core where most of the production of weakly-coupled ALPs takes place. Higher temperatures increase pion abundances in the SN core and the ALP production via pion conversion is significantly enhanced. Furthermore, ALP emission rates via $NN$ bremsstrahlung and pionic processes are both strongly dependent on the temperature [see Eq.~(4) and Eq.~(5) of Ref.~\cite{Lella:2022uwi}]. However, as well as for ALPs, also the neutrino emission is powered by higher temperatures and densities, so that the neutrino luminosity from the {\tt GARCHING} simulation at $t_{\mathrm{pb}}=1\,\s$ results to be 70\% higher than in the {\tt AGILE-BOLTZTRAN} simulation. This effect balances the enhancement of the ALP emissivity in the cooling criterion reported in Eq.~(\ref{eq:CoolingCondition}), leading to a difference of about 25\%. 
On the other hand, ALP emission rates in the trapping regime are not significantly affected by the change of the SN model and the change in the upper part of the cooling bound is essentially due to the different neutrino luminosity predicted by the two models. In this regime, the ALP luminosity decreases as $g_{ap}$ becomes larger, as shown in Fig.~\ref{fig:lavsgap}. Thus, the lower neutrino luminosity predicted by the {\tt AGILE-BOLTZTRAN} code implies that this model excludes slightly larger values of $g_{ap}$ in the trapping regime. Analogously, the KII event bound is not modified significantly, with the {\tt AGILE-BOLTZTRAN} code excluding a slightly larger region in the small-mass limit. To summarize, the two different SN models induce a $\sim 25-30\%$ uncertainty on the SN bounds, as shown by the hatched areas in the left panel of Fig.~\ref{fig:CoolingBound_withUncertainties}. \newline
The above discussion shows that different SN profiles (featuring different temperatures and densities) impact on the neutrino and the axion luminosity in a similar way. This implies that the SN cooling bound obtained by using the neutrino luminosity taken from the simulation (see Sec.~\ref{sec:ModifiedLum}) is not significantly affected. Therefore, we do not expect that even taking SN models predicting very high neutrino luminosities, such as those described in Ref.~\cite{Vartanyan:2023zlb}, the SN bounds would change dramatically.

\subsection{Presence of pions in the SN core}
The abundance of negatively charged pions in the SN core depends on the Equation of State (EoS) employed in the simulation and a self-consistent treatment of pionic matter in SNe is a topic still under investigation (see Ref.~\cite{Fore:2023gwv} for recent developments).
In order to estimate the maximum possible uncertainty coming from the abundance of pions in the SN core, we evaluate the relaxation of the SN bound in absence of pions, to obtain the most conservative estimate of the bound. In this scenario, the only possible production channel is $NN$ bremsstahlung and the cooling bound relaxes. The hatched blue region in the right panel of Fig.~\ref{fig:CoolingBound_withUncertainties} shows that the cooling bound in the free-streaming regime would be weakened by a factor $\sim 2$, excluding $g_{ap}\gtrsim 1.6\times 10^{-9}$ for $m_a < 10$~MeV, and it would probe masses $m_a \lesssim 150$~MeV. In particular, the constrained region of the parameter space results to be smaller at higher ALP masses, since bremsstrahlung is more affected by Boltzmann suppression than pionic processes. 
On the other hand, as discussed in Sec.~\ref{sec:ALPproduction}, pion conversions play a marginal role in the emission of ALPs from the SN core when entering the trapping regime. Therefore, the absence of pions in the PNS would have no impact on the SN cooling bound in the trapping regime and on the bound from the non-observation of extra events in the KII experiment.

\subsection{ALP absorption by heavy nuclei}
After the onset of the explosion, a large amount of energy is released by neutrinos during the Kelvin-Helmoltz cooling phase~(see Ref.~\cite{Mirizzi:2015eza} for an updated review). Meanwhile, the energy deposited by these neutrinos via capture and scattering events powers a baryonic outflow that expands with supersonic velocities, known as the neutrino-driven wind~\cite{Arnould:2007gh,Arcones:2012wj}. The presence of these nuclei at radii larger than $R_H\sim100\km$~\cite{Janka:2006fh} may represent an additional source of absorption for ALPs escaping the SN. In particular, species with atomic numbers similar to $^{16}$O, such as $^{9}$Be and $^{12}$C, and heavier nuclei, such as $^{56}$Fe, may be excited by ALP resonant absorption similarly to oxygen. This effect may lead to a significant reduction of the ALP flux reaching Earth for sufficiently high couplings. By assuming that the cross section for ALP absorption is of the same order of magnitude of the ALP-oxygen one, evaluated in Ref.~\cite{Carenza:2023wsm}, it is possible to estimate the reduction of the ALP flux.\\
The ALP absorption rate at each radius can be evaluated as
\begin{equation}
    \Gamma_H(E,r)\sim n_H(r)\,\sigma(E)\,,
\end{equation}
where $n_H(r)=\rho(r)\,X_H(r)/16\,m_N$ is the number density of heavy nuclei obtained by assuming that all the population is composed by species similar to oxygen \footnote{As shown in Ref.~\cite{Janka:2006fh}, at radii between 100 and 1000 $\km$, where the density is $\rho\sim10^5-10^6\,\g\cm^{-3}$, most of the nuclei are $^9$Be and $^{12}$C. At larger radii ($R\gtrsim10^4\,\km$) also heavier elements are present, but their contribution to the absorption is negligible, since in these regions the density is reduced to $\rho\sim10^2\,\g\cm^{-3}$.}, and $\rho(r)$ and $X_H(r)$ are the matter density and the heavy nuclei fraction taken from the SN profile, respectively. Then, the flux of ALPs escaping the SN with a given energy $E$ is reduced by a fraction
\begin{equation}
    \eta_H(E)=\exp{\left[-\int_{R_H}^\infty\Gamma_H(E,r)\,dr\right]}\,.
\end{equation}
The green hatched region in Fig.~\ref{fig:CoolingBound_withUncertainties} displays the modification of the bound by taking into account this effect. It is remarkable to notice that at $g_{ap}=3\times10^{-5}$ the ALP flux is reduced just by 20\%. Therefore, the KII event bound is significantly affected by absorption effects only in a region of the parameter space which is already excluded by the SNO bound. Comparing the two panels in Fig.~\ref{fig:CoolingBound_withUncertainties}, one can conclude that the uncertainties induced by the SN models are subdominant compared to the ones related to the presence of pions in the SN core and the possible ALP absorption by heavy nuclei. Thus, the right panel of Fig.~\ref{fig:CoolingBound_withUncertainties} is a good estimator of the total uncertainty affecting the SN bounds computed in this work.

\section{SN bounds on the QCD axion mass}
\label{sec:BoundQCDaxion}
\begin{figure}[t!]
    \includegraphics[width=1\columnwidth]{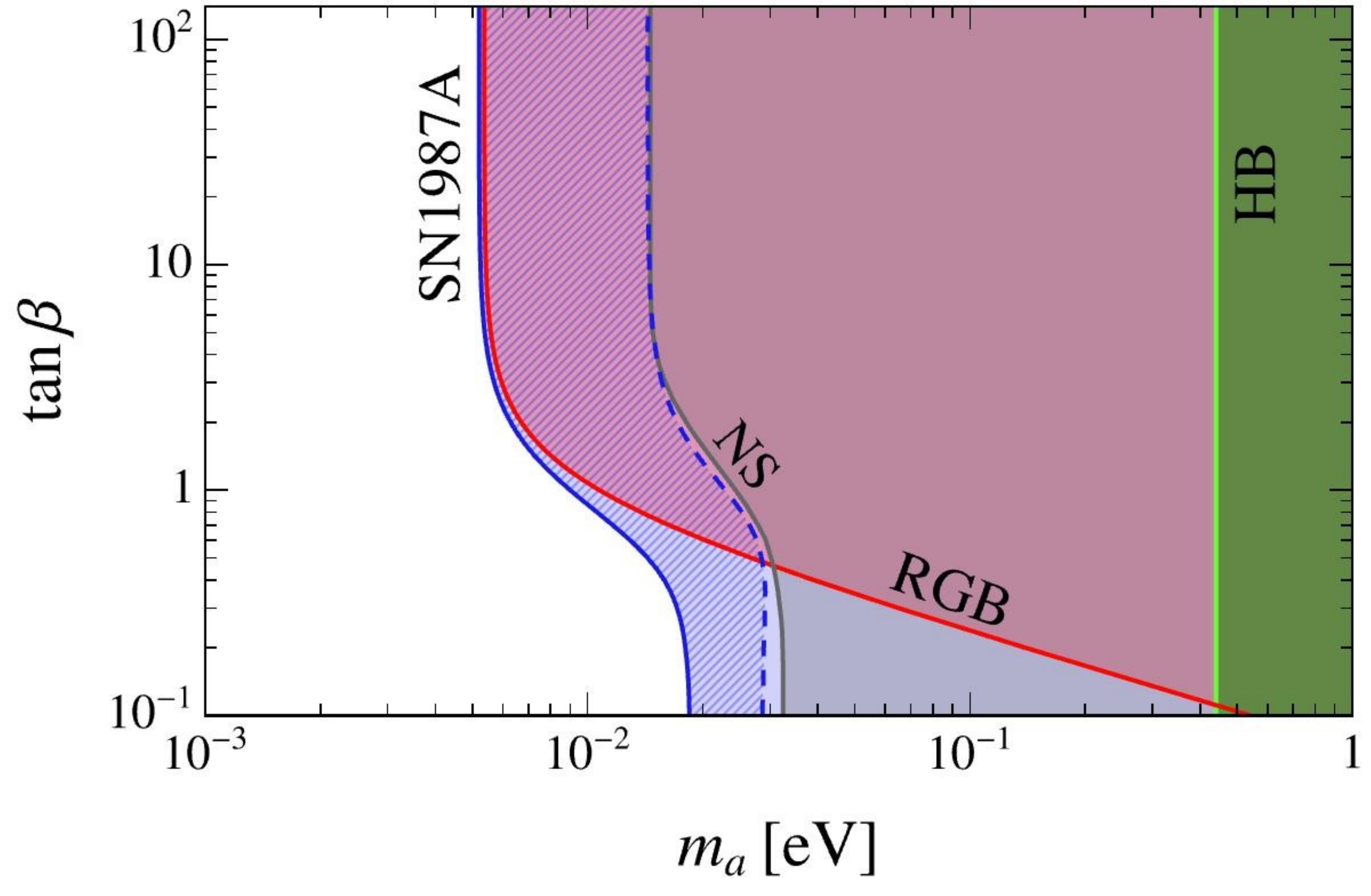}
    \caption{Summary plot of the bounds in the $\tan\beta$ {\emph vs} $\$,m_a$ plane for the DFSZ axion model. The blue shadowed region, delimited by the blue solid line, displays the region excluded by the SN bounds placed in this work. The hatched blue region between the solid and the dashed blue lines represents uncertainties related to the presence of pions. The red shadowed area shows the region of the parameter space ruled out by the RGB tips bound~\cite{Capozzi:2020cbu}, while the grey and green shadowed region refer to the bound placed by isolated neutron star cooling~\cite{Buschmann:2021juv} and the duration of the HB phase in Globular Clusters~\cite{Carenza:2020zil}.}
    \label{fig:ma_vs_tanbeta}
\end{figure}

In the case of the canonical QCD axion, the SN bounds placed in this work can be read in terms of constraints on the QCD axion mass. In particular, given the axion coupling to nucleons $g_{aN}$, the corresponding QCD axion mass is obtained as~\cite{ParticleDataGroup:2022pth}
\begin{equation}
    m_a\simeq 6.06\,\left(\frac{g_{aN}}{10^{-6}\,C_{aN}}\right)\,\eV\,.
\end{equation}
In the case of the KSVZ axion model, which we assumed as benchmark model for our analysis, the SN mass bound can be read directly from Fig.~\ref{fig:CoolingBound_withUncertainties}, where the solid dark-yellow line and the shadowed yellow band refer to the KSVZ and Dine-Fischler-Srednicki-Zhitnitsky (DFSZ) axion models, respectively~\cite{GrillidiCortona:2015jxo}.
One realizes that SN arguments place a bound on the KSVZ axion masses that varies in the range
\begin{equation}
    m_a\in[8,19]\,\meV\,,
\end{equation}
depending on the presence of pions in the SN core. These results, within the uncertainties discussed in the previous sections, are comparable with the bound placed in Ref.~\cite{Buschmann:2021juv}, which ruled out KSVZ axion masses ${m_a\gtrsim16\,\meV}$ from isolated neutron star cooling. We highlight that in the KSVZ axion model also the axion-photon coupling $g_{a\gamma}$ is switched on. However, the SN bounds evaluated in this work are remarkably stronger than constraints from the horizontal branch (HB) stars in Globular Cluster~\cite{Ayala:2014pea,Carenza:2020zil}, excluding ${g_{a\gamma}\gtrsim0.66\times10^{-10}\,\GeV^{-1}}$, which corresponds to $m_a\gtrsim440\,\meV$~\cite{ParticleDataGroup:2022pth}.
\newline
Our results can be simply extended also to other canonical QCD axion models by substituting different values for the model dependent constants $C_{aN}$. In particular, in the DFSZ axion model one has~\cite{Dine:1981rt,Zhitnitsky:1980tq}
\begin{equation}
    \begin{split}
        C_{ap}&=−0.435\sin^2{\beta}-0.182\pm 0.025\,\\
        C_{an}&=-0.414 \sin^2{\beta}-0.16\pm 0.025\,,
    \end{split}
\end{equation}
with $\tan\beta$ confined to $0.28<\tan{\beta}< 140$ by perturbative unitarity constraints~\cite{GrillidiCortona:2015jxo,DiLuzio:2020wdo} (see Ref.~\cite{DiLuzio:2023tqe} for a more recent evaluation
of the DFSZ axion-proton coupling range). The solid blue line in Fig.~\ref{fig:ma_vs_tanbeta} displays the constraint on the DFSZ axion mass placed in this work as a function of $\tan\beta$. In particular, SN arguments exclude $m_a\gtrsim5 \meV$ for $\tan\beta\gtrsim5$ and $m_a\gtrsim18\,\meV$ for $\tan\beta\lesssim1$. In the absence of pions this bound is relaxed to $m_a\gtrsim14 \meV$ for $\tan\beta\gtrsim5$ and $m_a\gtrsim28\,\meV$ for $\tan\beta\lesssim1$ (see the dashed blue line). Moreover, since in DFSZ axion models also coupling to leptons are available, it is necessary to take into account also the red giant branch (RGB) tip bound introduced in Refs.~\cite{Straniero:2020iyi,Capozzi:2020cbu}. It is remarkable to notice that, in presence of pions, SN 1987A bounds are comparable to the RGB constraint, as shown by the red line in Fig.~\ref{fig:ma_vs_tanbeta}, representing the bound from Ref.~\cite{Capozzi:2020cbu}.

\begin{table}[t!]
\caption{SN luminosity of QCD axion compared with neutrino luminosity at $t_{\rm pb}=1$ s for different axion masses, where $m_{a}=100$~meV roughly corresponds to the reach of cosmological searches for hot dark matter axions~\cite{Archidiacono:2015mda}.
}
 \centering
\begin{tabular}{|c|c|}
\hline
$m_a$ [meV] &\,\,$L_a/L_\nu$\\
\hline
\hline
$1$ &   0.016 \\
$10$ &   1.56 \\
$100$ & 67.5 \\
\hline
\end{tabular}
 \label{tab:Luminosity}
\end{table}

\section{Discussion and Conclusions} 
\label{sec:Conclusions}
In this work, we have set stringent SN bounds on ALPs coupled to nucleons. In particular, we have extended the bounds on free-streaming ALPs to the case of strongly-coupled ALPs, updating the results in the recent literature, and constrained them from the non-observation of ALP-induced events in KII.
The combination of these limits allows us to exclude values of the ALP-proton coupling $g_{ap}\gtrsim6\times10^{-10}$ for ALP masses $m_a\lesssim1\,\MeV$. 
Therefore, SN bounds strongly constrain the parameter space available for ALPs coupled to nucleons.

This is in contrast with the original literature on the SN 1987A bound, which reported the existence of a window around a QCD axion mass $m_a \sim {\mathcal O}(1)$ eV which was not excluded by the energy-loss argument nor by the axion events in the KII (see, e.g.,~\cite{Raffelt:1996wa} for a detailed discussion). 
This region was  dubbed the ``hadronic axion window''~\cite{Moroi:1998qs,Chang:1993gm}. 
It was later shown that cosmological mass bounds on axions would close this window~\cite{Hannestad:2005df}. 
Nevertheless, in non-standard cosmologies, one can relax the axion mass bounds~\cite{Grin:2007yg}, reopening this region of the parameter space. 
In this regard, it has been recently shown that in low-reheating scenarios multi-eV axions are allowed by cosmology~\cite{Carenza:2021ebx}. 
Therefore, the SN 1987A argument is extremely useful to give an independent constraint:
with our new analysis we show that SN bounds are enough to exclude canonical QCD axion models  (reported in the yellow band in Fig.~\ref{fig:CoolingBound}) for masses $m_a \gtrsim \mathcal{O}(10)\,\meV$. 
The absence of pions in the SN core, whose abundance is still debated, would relax the aforementioned constraint by a factor $\sim2$.
This bound is comparable with the recent one placed from the cooling of young neutron stars~\cite{Buschmann:2021juv} and it is stronger than the reach of current and future cosmological experiments, which would probe axion masses $m_a\gtrsim150  \meV$~\cite{Archidiacono:2015mda,DEramo:2022nvb}, as depicted by the vertical orange line on the QCD axion band in Fig.~\ref{fig:CoolingBound}.  
Note that for $m_a\sim {\mathcal O}(100)  \meV$, QCD axions in SNe would be in an intermediate regime between the complete free-streaming case and the strongly-trapped one.  
As shown in 
Table~\ref{tab:Luminosity} for the axion mass range probed by cosmology, the SN axion luminosity at $t_{\text{pb}}=1\,\s$ would 
be significantly higher than the neutrino one, implying that the axion emissivity would dominate the SN cooling. This would be in tension even with the early time neutrino signal of SN~1987A. 
Therefore, contrarily to the neutrino case, it is unlikely that future cosmological probes would find signatures of the QCD axion mass as hot dark matter. Nevertheless, below the SN bound, QCD axions with masses $m_a\lesssim 10\,\meV$ may contribute to subleading cosmological dark radiation~\cite{DEramo:2022nvb}, which is in the reach of future cosmological surveys (see Ref.~\cite{Baumann:2016wac}).

\acknowledgments
We warmly thank  Thomas Janka for giving us access to the {\tt GARCHING} group archive. PC thanks Benjamin Wallisch for useful discussions.
This article is based upon work from COST Action COSMIC WISPers CA21106, supported by COST (European Cooperation in Science and Technology).
This work is (partially) supported
by ICSC – Centro Nazionale di Ricerca in High Performance Computing,
 Big Data and Quantum Computing, funded by European Union - NextGenerationEU. The work of A.L.,  G.L., A.M., 
was partially supported by the research grant number 2017W4HA7S ``NAT-NET: Neutrino and Astroparticle Theory Network'' under the program PRIN 2017 funded by the Italian Ministero dell'Università e della Ricerca (MUR).
The work of A.L. and A.M.  was also partially supported by the research grant number 2022E2J4RK "PANTHEON: Perspectives in Astroparticle and
Neutrino THEory with Old and New messengers" under the program PRIN 2022 funded by the Italian Ministero dell’Universit\`a e della Ricerca (MUR).
The work of P.C. is supported by the European Research Council under Grant No.~742104 and by the Swedish Research Council (VR) under grants  2018-03641 and 2019-02337. 
T.R. was partially supported by the COST action ChETEC (CA16117). P.C., M.G. and G.L. thank the Galileo Galilei Institute for Theoretical Physics for 
hospitality during the preparation of part of this work.

\bibliographystyle{bibi.bst}
\bibliography{references.bib}

\vspace{1 cm}

\clearpage

\setcounter{equation}{0}
\setcounter{figure}{0}
\setcounter{table}{0}
\setcounter{section}{0}
\makeatletter
\renewcommand{\theequation}{A\arabic{equation}}
\renewcommand{\thefigure}{A\arabic{figure}}
\renewcommand{\thetable}{A\arabic{table}}

\onecolumngrid

\appendix
\section{Heavy ALPs coupled with nucleons in the trapping regime}
\label{app:trapping}

\subsection{ALP absorption in a nuclear medium}
\label{app:absorption}
Interactions between ALPs and nuclear matter are described by the following effective Lagrangian~\cite{DiLuzio:2020wdo,Choi:2021ign,Ho:2022oaw}:

\begin{equation}
    \begin{split}
        \mathcal{L}_{\rm{int}}&=\frac{g_A}{2f_\pi}\left[\partial_\mu\pi^0(\Bar{p}\gamma^\mu\gamma_5p-\Bar{n}\gamma^\mu\gamma_5n)+\sqrt{2}\partial_\mu\pi^+\Bar{p}\gamma^\mu\gamma_5n+\sqrt{2}\partial_\mu\pi^-\Bar{n}\gamma^\mu\gamma_5p\right]\\
        &+g_a\frac{\partial^\mu a}{2m_N}\left[C_{ap}\Bar{p}\gamma^\mu\gamma_5p+C_{an}\Bar{n}\gamma^\mu\gamma_5n+\frac{C_{a\pi N}}{f_\pi}(i\pi^+\Bar{p}\gamma^\mu n-i\pi^-\Bar{n}\gamma^\mu p)\right]\\
        &+\frac{\mathcal{C}}{\sqrt{6} f_\pi}\left(\Bar{n}\,\Delta^+_\mu\partial^\mu\pi^-+\overline{\Delta^+_\mu}\,n\,\partial^\mu\pi^+
        -\Bar{p}\,\Delta^0_\mu\partial^\mu\pi^+
        -\overline{\Delta^0_\mu}\,p\,\partial^\mu\pi^-\right)\\
        &+g_a\frac{\partial^\mu a}{2m_N}C_{aN\Delta}\left[\Bar{p}\,\Delta^+_\mu+\overline{\Delta^+_\mu}\,p+\Bar{n}\,\Delta^0_\mu+\overline{\Delta^0_\mu}\,n\right],
    \end{split}
\label{eq:NuclearInteractions}
\end{equation}

where $g_a$ encodes the strength of the ALP couplings with nucleons, $f_{\pi}=92.4~\MeV$ is the pion decay constant, $C_{a\pi N}=(C_{ap}-C_{an})/\sqrt{2}\,g_{A}$~\cite{Choi:2021ign} and ${C_{aN\Delta}=-\sqrt{3}/2\,(C_{ap}-C_{an})}$~\cite{Ho:2022oaw}, while $g_{A}=1.28$ is the axial coupling. 
In particular, the first and the third lines describe nucleons interactions with pions and the $\Delta-$resonance, while the second and the fourth lines contain ALPs couplings.
Notice that this Lagrangian includes the contribution coming from the contact interaction term, which was originally discussed in Ref.~\cite{Carena:1988kr} and recently re-discussed in Ref.~\cite{Choi:2021ign}. Moreover, we have included vertices for ALP-$\Delta$ interactions which have been recently introduced in Ref.~\cite{Ho:2022oaw}, where it has been estimated that they could give observable contributions to the ALP emission from a SN core.
It is convenient to define the ALP couplings with protons and neutrons as $g_{aN}=g_a\, C_{aN}$, for $N=p,n$, where $C_{aN}$ are model-dependent coupling constant. In this work we will use as benchmark values for the model-dependent constants $C_{ap}=-0.47$ and $C_{an}=0$, inspired by the Kim-Shifman-Vainshtein-Zakharov (KSVZ) axion model~\cite{GrillidiCortona:2015jxo}. Our results can be simply extended also to other models just by substituting different values for these constants. As an example in the Dine-Fischler-Srednicki-Zhitnitsky (DFSZ) axion model~\cite{Dine:1981rt,Zhitnitsky:1980tq}
\begin{equation}
    \begin{split}
        C_{ap}&=−0.617 + 0.435 \sin^2{\beta} \pm 0.025\,,\\
        C_{an}&= 0.254 - 0.414 \sin^2{\beta} \pm 0.025\,,
    \end{split}
\end{equation}
with $\tan\beta$ confined to $0.28<\tan{\beta}< 140$ by perturbative unitarity constraints~\cite{GrillidiCortona:2015jxo,DiLuzio:2020wdo} (see Ref.~\cite{DiLuzio:2023tqe} for a more recent evaluation
of the DFSZ axion-proton coupling range). 
As discussed in the main text (see also Ref.~\cite{Lella:2022uwi}), ALPs can be produced by means of $NN$ bremsstrahlung~\cite{Carena:1988kr,Brinkmann:1988vi,Raffelt:1993ix,Raffelt:1996wa,Carenza:2019pxu} and pionic Compton-like processes~\cite{Raffelt:1993ix,Keil:1996ju,Carenza:2020cis}.
\newline
Once produced, ALPs could be reabsorbed in the nuclear medium inside the SN core just by means of reverse processes
\begin{equation}
    \begin{split}
        N+N+a&\rightarrow N+N\\
        N+a&\rightarrow N+\pi\,. 
    \end{split}
\end{equation}

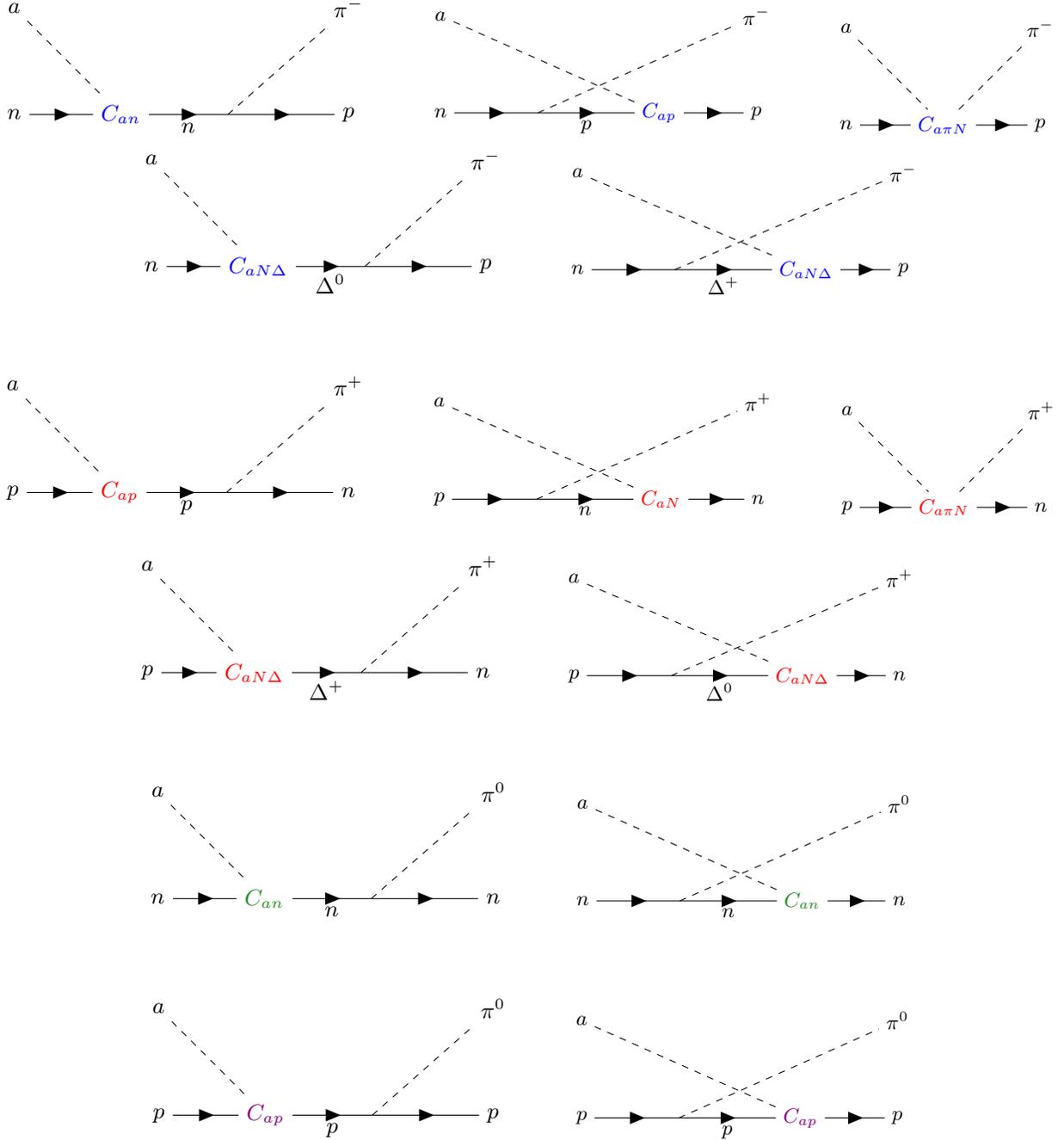
\begin{figure*}[!h]
\centering
\makebox[14 cm] [c] {
\begin{tikzpicture}[scale=1.1, transform shape]
\begin{feynman}
\vertex (a) {\(n\)} ;
\vertex [above=of a] (b) {\(a\)};
\vertex [right=of a] (c) {\(\textcolor{blue}{C_{an}}\)};
\vertex [right=of c] (d);
\vertex [right=of d] (e) {\(p\)};
\vertex [above=of e] (f) {\(\pi^-\)};

\diagram* {
(a) -- [fermion] (c),
(c) -- [fermion, edge label'=\(n\)] (d),
(d) -- [fermion] (e),
(b) -- [scalar] (c),
(d) -- [scalar] (f),
};
\end{feynman}
\end{tikzpicture}

\hspace{0.65cm}

\begin{tikzpicture}[scale=1., transform shape]
\begin{feynman}
\vertex (a) {\(n\)} ;
\vertex [above=of a] (b) {\(a\)};
\vertex [right=of a] (c);
\vertex [right=of c] (d) {\(\textcolor{blue}{C_{ap}}\)};
\vertex [right=of d] (e) {\(p\)};
\vertex [above=of e] (f) {\(\pi^-\)};

\diagram* {
(a) -- [fermion] (c),
(c) -- [fermion, edge label'=\(p\)] (d),
(d) -- [fermion] (e),
(b) -- [scalar] (d),
(c) -- [scalar] (f),
};
\end{feynman}
\end{tikzpicture}

\hspace{0.65cm}

\begin{tikzpicture}[scale=1., transform shape]
\begin{feynman}
\vertex (a) {\(n\)} ;
\vertex [above=of a] (b) {\(a\)};
\vertex [right=of a] (c) {\(\textcolor{blue}{C_{a\pi N}}\)};
\vertex [right=of c] (d) {\(p\)};
\vertex [above=of d] (e) {\(\pi^-\)};

\diagram* {
(a) -- [fermion] (c),
(c) -- [fermion] (d),
(b) -- [scalar] (c),
(c) -- [scalar] (e),
};
\end{feynman}
\end{tikzpicture}
}

\makebox[14 cm] [c] {
\begin{tikzpicture}[scale=1.1, transform shape]
\begin{feynman}
\vertex (a) {\(n\)} ;
\vertex [above=of a] (b) {\(a\)};
\vertex [right=of a] (c) {\(\textcolor{blue}{C_{aN\Delta}}\)};
\vertex [right=of c] (d);
\vertex [right=of d] (e) {\(p\)};
\vertex [above=of e] (f) {\(\pi^-\)};

\diagram* {
(a) -- [fermion] (c),
(c) -- [fermion, edge label'=\(\Delta^0\)] (d),
(d) -- [fermion] (e),
(b) -- [scalar] (c),
(d) -- [scalar] (f),
};
\end{feynman}
\end{tikzpicture}

\hspace{0.65cm}

\begin{tikzpicture}[scale=1., transform shape]
\begin{feynman}
\vertex (a) {\(n\)} ;
\vertex [above=of a] (b) {\(a\)};
\vertex [right=of a] (c);
\vertex [right=of c] (d) {\(\textcolor{blue}{C_{aN\Delta}}\)};
\vertex [right=of d] (e) {\(p\)};
\vertex [above=of e] (f) {\(\pi^-\)};

\diagram* {
(a) -- [fermion] (c),
(c) -- [fermion, edge label'=\(\Delta^+\)] (d),
(d) -- [fermion] (e),
(b) -- [scalar] (d),
(c) -- [scalar] (f),
};
\end{feynman}
\end{tikzpicture}
}

\vspace{1cm}

\centering
\makebox[14 cm] [c] {
\begin{tikzpicture}[scale=1.1, transform shape]
\begin{feynman}
\vertex (a) {\(p\)} ;
\vertex [above=of a] (b) {\(a\)};
\vertex [right=of a] (c) {\(\textcolor{red}{C_{ap}}\)};
\vertex [right=of c] (d);
\vertex [right=of d] (e) {\(n\)};
\vertex [above=of e] (f) {\(\pi^+\)};

\diagram* {
(a) -- [fermion] (c),
(c) -- [fermion, edge label'=\(p\)] (d),
(d) -- [fermion] (e),
(b) -- [scalar] (c),
(d) -- [scalar] (f),
};
\end{feynman}
\end{tikzpicture}

\hspace{0.65cm}

\begin{tikzpicture}[scale=1., transform shape]
\begin{feynman}
\vertex (a) {\(p\)} ;
\vertex [above=of a] (b) {\(a\)};
\vertex [right=of a] (c);
\vertex [right=of c] (d) {\(\textcolor{red}{C_{aN}}\)};
\vertex [right=of d] (e) {\(n\)};
\vertex [above=of e] (f) {\(\pi^+\)};

\diagram* {
(a) -- [fermion] (c),
(c) -- [fermion, edge label'=\(n\)] (d),
(d) -- [fermion] (e),
(b) -- [scalar] (d),
(c) -- [scalar] (f),
};
\end{feynman}
\end{tikzpicture}

\hspace{0.65cm}

\begin{tikzpicture}[scale=1., transform shape]
\begin{feynman}
\vertex (a) {\(p\)} ;
\vertex [above=of a] (b) {\(a\)};
\vertex [right=of a] (c) {\(\textcolor{red}{C_{a\pi N}}\)};
\vertex [right=of c] (d) {\(n\)};
\vertex [above=of d] (e) {\(\pi^+\)};

\diagram* {
(a) -- [fermion] (c),
(c) -- [fermion] (d),
(b) -- [scalar] (c),
(c) -- [scalar] (e),
};
\end{feynman}
\end{tikzpicture}
}

\makebox[14 cm] [h!] {
\begin{tikzpicture}[scale=1.1, transform shape]
\begin{feynman}
\vertex (a) {\(p\)} ;
\vertex [above=of a] (b) {\(a\)};
\vertex [right=of a] (c) {\(\textcolor{red}{C_{ap}}\)};
\vertex [right=of c] (d);
\vertex [right=of d] (e) {\(n\)};
\vertex [above=of e] (f) {\(\pi^+\)};

\diagram* {
(a) -- [fermion] (c),
(c) -- [fermion, edge label'=\(n\)] (d),
(d) -- [fermion] (e),
(b) -- [scalar] (c),
(d) -- [scalar] (f),
};
\end{feynman}
\end{tikzpicture}

\hspace{0.65cm}

\begin{tikzpicture}[scale=1., transform shape]
\begin{feynman}
\vertex (a) {\(p\)} ;
\vertex [above=of a] (b) {\(a\)};
\vertex [right=of a] (c);
\vertex [right=of c] (d) {\(\textcolor{red}{C_{an}}\)};
\vertex [right=of d] (e) {\(n\)};
\vertex [above=of e] (f) {\(\pi^+\)};

\diagram* {
(a) -- [fermion] (c),
(c) -- [fermion, edge label'=\(p\)] (d),
(d) -- [fermion] (e),
(b) -- [scalar] (d),
(c) -- [scalar] (f),
};
\end{feynman}
\end{tikzpicture}

\hspace{0.65cm}

\begin{tikzpicture}[scale=1., transform shape]
\begin{feynman}
\vertex (a) {\(p\)} ;
\vertex [above=of a] (b) {\(a\)};
\vertex [right=of a] (c) {\(\textcolor{red}{C_{a\pi N}}\)};
\vertex [right=of c] (d) {\(n\)};
\vertex [above=of d] (e) {\(\pi^+\)};

\diagram* {
(a) -- [fermion] (c),
(c) -- [fermion] (d),
(b) -- [scalar] (c),
(c) -- [scalar] (e),
};
\end{feynman}
\end{tikzpicture}
}

\makebox[14 cm] [c] {
\begin{tikzpicture}[scale=1.1, transform shape]
\begin{feynman}
\vertex (a) {\(p\)} ;
\vertex [above=of a] (b) {\(a\)};
\vertex [right=of a] (c) {\(\textcolor{red}{C_{aN\Delta}}\)};
\vertex [right=of c] (d);
\vertex [right=of d] (e) {\(n\)};
\vertex [above=of e] (f) {\(\pi^+\)};

\diagram* {
(a) -- [fermion] (c),
(c) -- [fermion, edge label'=\(\Delta^+\)] (d),
(d) -- [fermion] (e),
(b) -- [scalar] (c),
(d) -- [scalar] (f),
};
\end{feynman}
\end{tikzpicture}

\hspace{0.65cm}

\begin{tikzpicture}[scale=1., transform shape]
\begin{feynman}
\vertex (a) {\(p\)} ;
\vertex [above=of a] (b) {\(a\)};
\vertex [right=of a] (c);
\vertex [right=of c] (d) {\(\textcolor{red}{C_{aN\Delta}}\)};
\vertex [right=of d] (e) {\(n\)};
\vertex [above=of e] (f) {\(\pi^+\)};

\diagram* {
(a) -- [fermion] (c),
(c) -- [fermion, edge label'=\(\Delta^0\)] (d),
(d) -- [fermion] (e),
(b) -- [scalar] (d),
(c) -- [scalar] (f),
};
\end{feynman}
\end{tikzpicture}
}
\vspace{1cm}

\centering
\makebox[14 cm] [c] {
\begin{tikzpicture}[scale=1.1, transform shape]
\begin{feynman}
\vertex (a) {\(n\)} ;
\vertex [above=of a] (b) {\(a\)};
\vertex [right=of a] (c) {\(\textcolor{ForestGreen}{C_{an}}\)};
\vertex [right=of c] (d);
\vertex [right=of d] (e) {\(n\)};
\vertex [above=of e] (f) {\(\pi^0\)};

\diagram* {
(a) -- [fermion] (c),
(c) -- [fermion, edge label'=\(n\)] (d),
(d) -- [fermion] (e),
(b) -- [scalar] (c),
(d) -- [scalar] (f),
};
\end{feynman}
\end{tikzpicture}

\hspace{0.65cm}

\begin{tikzpicture}[scale=1., transform shape]
\begin{feynman}
\vertex (a) {\(n\)} ;
\vertex [above=of a] (b) {\(a\)};
\vertex [right=of a] (c);
\vertex [right=of c] (d) {\(\textcolor{ForestGreen}{C_{an}}\)};
\vertex [right=of d] (e) {\(n\)};
\vertex [above=of e] (f) {\(\pi^0\)};

\diagram* {
(a) -- [fermion] (c),
(c) -- [fermion, edge label'=\(n\)] (d),
(d) -- [fermion] (e),
(b) -- [scalar] (d),
(c) -- [scalar] (f),
};
\end{feynman}
\end{tikzpicture}
}

\vspace{1cm}

\centering
\makebox[14 cm] [c] {
\begin{tikzpicture}[scale=1.1, transform shape]
\begin{feynman}
\vertex (a) {\(p\)} ;
\vertex [above=of a] (b) {\(a\)};
\vertex [right=of a] (c) {\(\textcolor{violet}{C_{ap}}\)};
\vertex [right=of c] (d);
\vertex [right=of d] (e) {\(p\)};
\vertex [above=of e] (f) {\(\pi^0\)};

\diagram* {
(a) -- [fermion] (c),
(c) -- [fermion, edge label'=\(p\)] (d),
(d) -- [fermion] (e),
(b) -- [scalar] (c),
(d) -- [scalar] (f),
};
\end{feynman}
\end{tikzpicture}

\hspace{0.65cm}

\begin{tikzpicture}[scale=1., transform shape]
\begin{feynman}
\vertex (a) {\(p\)} ;
\vertex [above=of a] (b) {\(a\)};
\vertex [right=of a] (c);
\vertex [right=of c] (d) {\(\textcolor{violet}{C_{ap}}\)};
\vertex [right=of d] (e) {\(p\)};
\vertex [above=of e] (f) {\(\pi^0\)};

\diagram* {
(a) -- [fermion] (c),
(c) -- [fermion, edge label'=\(p\)] (d),
(d) -- [fermion] (e),
(b) -- [scalar] (d),
(c) -- [scalar] (f),
};
\end{feynman}
\end{tikzpicture}
}
\caption{Feynman diagrams of pionic Compton processes for ALP absorption, including also the contact and the delta mediated diagrams. The upper diagrams refer to $a+n\rightarrow \pi^-+p$, the middle ones depicts $a+p\rightarrow \pi^++n$, while the lower ones refer to $a+n\rightarrow \pi^0+n$.}
\label{fig:PionConvDiagrams}
\end{figure*}

In the previous literature~\cite{Carenza:2020cis,Choi:2021ign} it has been argued that only processes involving negatively charged pions are relevant for ALP production. This is due to the fact that the abundance of $\pi^+$ and $\pi^0$ inside a SN core is strongly suppressed with respect to $\pi^-$, as recently stressed in Ref.~\cite{Fore:2019wib}. 
However, in axion absorption via pionic processes, pions are present only in the final state. Thus, the concentration of the different species does not influence the absorption rates. In this regard, we remark that the pion abundance $Y_\pi$ is just involved in the computation of the chemical potentials $\mu_\pi$ contained in the Bose stimulating factors $1+f_\pi\sim1$. As a consequence, it is necessary to include the absorption contributions given by all the possible pionic processes
\begin{equation}
    \begin{split}
    &\color{blue}{n+a\rightarrow p+\pi^-}\\
    &\color{red}{p+a\rightarrow n+\pi^+}\\
    &\color{ForestGreen}{n+a\rightarrow n+\pi^0}\\
    &\color{violet}{p+a\rightarrow p+\pi^0}\,.
    \end{split}
\end{equation}

Referring to Fig.~\ref{fig:PionConvDiagrams} and to the interaction Lagrangian introduced in Eq.~\eqref{eq:NuclearInteractions}, the scattering amplitudes for other processes can be easily computed from the matrix element for ${n+a\rightarrow p+\pi^-}$ following some simple prescriptions: 
\begin{itemize}
    \item To switch from $a+n\rightarrow p+\pi^-$ to $a+p\rightarrow n+\pi^+$ it is only needed to exchange $C_{ap}\leftrightarrow C_{an}$.
    
    \item The relative sign between the Compton diagrams and the contact interaction term is different in $a+p\rightarrow n+\pi^+$ with respect to $a+n\rightarrow p+\pi^-$. 
    However, this is not relevant for us since the interference terms involving the contact diagram have been neglected in this work, being of higher order in $1/m_N$ (see Refs.~\cite{Carena:1988kr,Choi:2021ign,Ho:2022oaw} for further details).
    
    \item In processes involving neutral pions only Compton-like diagrams are possible. Moreover, it is necessary to divide by a factor of 2 with respect to the same contribution estimated for charged pions~\cite{bjorken1965relativistic}.
    
    \item The integral on the nucleons kinetic energies in the formulas in Eq.~(5) of  Ref.~\cite{Lella:2022uwi} has to be generalized as follows
        \begin{equation}
            \int_0^\infty dy\, y^2\frac{1}{\exp{\left(y^2-\hat{\mu}_{\text{in}}\right)}+1}\frac{1}{\exp{\left(-y^2+\hat{\mu}_{\text{out}}\right)}+1}\,,
        \end{equation}
    where $\hat{\mu}_{\text{in}}$ is the relativistic chemical potential of the incoming nucleon while $\hat{\mu}_{\text{out}}$ refers to the outgoing nucleon.

\end{itemize}

\begin{figure} [t!]
\centering
\includegraphics[scale=0.8]{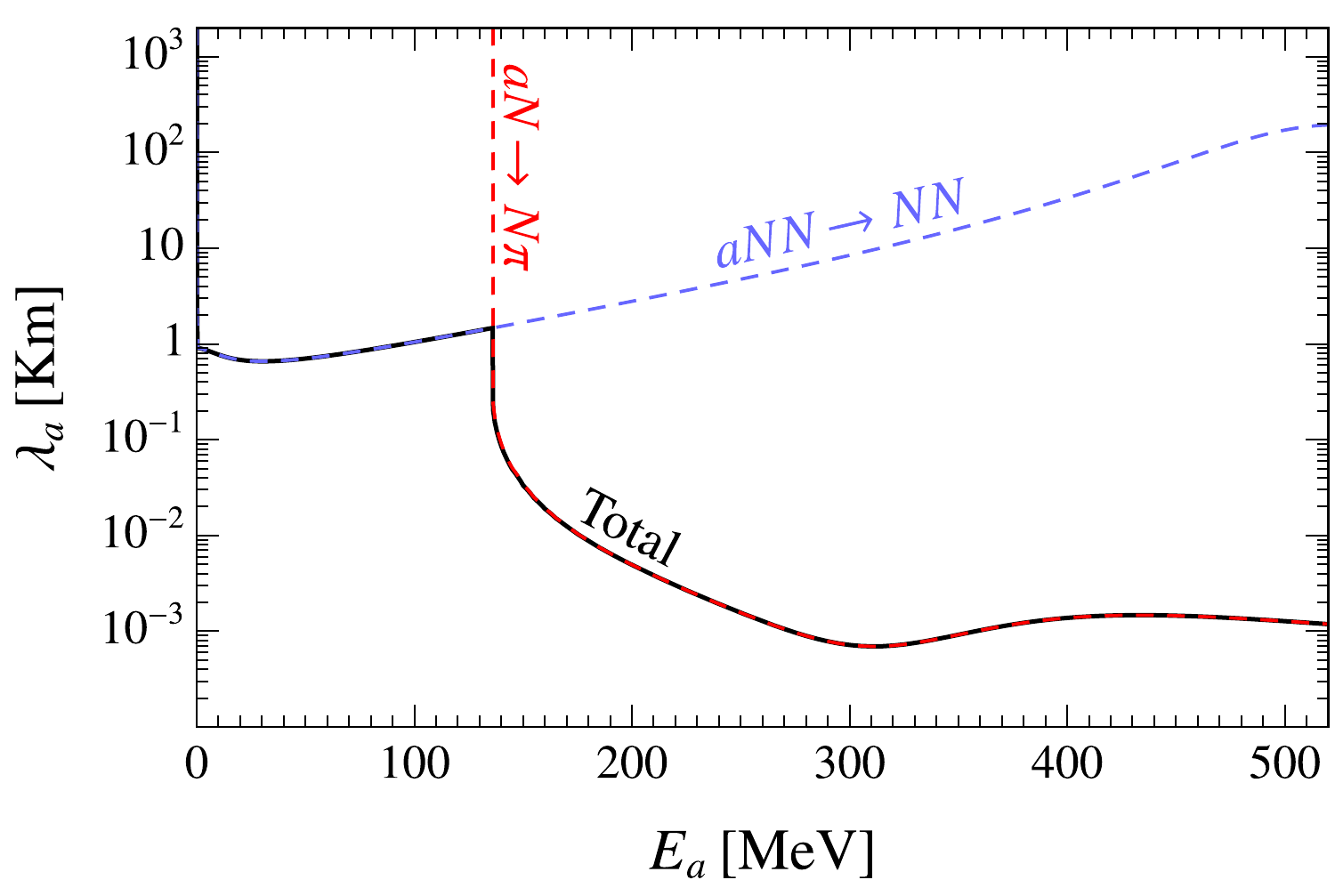}

\caption{Behaviour of the ALP mean free path at the value of the upper bound  for the case of massless ALPs at $R\simeq20\,\km$. In particular the dashed blue line describes the absorption length by means of inverse bremsstrahlung, the red one by means of inverse pion conversion and the black solid curve shows the total MFP.}
	\label{fig:MFP}
\end{figure}

\subsection{Mean free path}
\label{app:mfp}
Starting from the expressions for the ALP spectrum of production $d^2n_a/dE_a\,dt$ introduced in Ref.~\cite{Lella:2022uwi}, the mean free path (MFP) $\lambda_a$ associated with the inverse pion conversion and inverse bremsstrahlung can be computed as~\cite{Giannotti:2005tn}
\begin{equation}
    \begin{split}
        \lambda_a^{-1}(E_a)&=\frac{1}{2|\textbf{p}_a|}\,\frac{d^2 n_a (\chi\,E_a)}{d\Pi_a\,dt}=\\
        &2\pi^2\,(E_a^2-m_a^2)^{-1}\,\frac{d^2 n_a (\chi\,E_a)}{dE_a\,dt}\,,
    \end{split}
\end{equation}
where $E_a$ and $p_a$ are the ALP energy and four-momentum, $d\Pi_a=d^3\textbf{p}_a/(2\pi)^3 2E_a$ is the ALP phase space 
and $\chi=\pm1$ for pionic processes and bremsstrahlung respectively. This change in sign of the ALP energy is due to the different role that the ALP plays in the two processes. In bremsstrahlung processes, depending on whether the ALP is absorbed or emitted, its energy is released to or soaked up from the nuclear medium and then it must change sign switching from the production to the absorption rate (see, e.g.~\cite{Raffelt:1996wa}). On the other hand, in pionic processes the conservation of the energy for non-relativistic nucleons just requires that the ALP and the pion involved must have the same energy and no change in sign is necessary. However, note that in pionic absorption the pion is present in the final state implying the substitution $f_\pi\rightarrow1+f_\pi$.
Let us highlight that the expression of the MFP for inverse bremsstrahlung coincides with Eq.~(4.27) in~\cite{Raffelt:1996wa} extended to the case of massive ALPs. In particular we have used the expression of the structure functions $S_\sigma=\Gamma_\sigma/(E_a^2+\Gamma^2)\,s_\sigma$ in a generic degeneracy regime given in Ref.~\cite{Carenza:2019pxu}, where the integral functions in $s_\sigma$ have to be evaluated with the substitution $x\rightarrow-x$ and $\Gamma$ has to be chosen in order to have properly normalized structure functions~\cite{Sawyer:1989nu}.\newline 
The behaviour of the MFP in the massless case $m_a<10\,\MeV$ at $R\simeq16.5\,\km$ is depicted in Fig.~\ref{fig:MFP}. In inverse bremsstrahlung the incoming ALP just yields a certain amount of kinetic energy to the system of nucleons. As a consequence, ALPs absorption is favored (shorter MFP) when they have energies of the order of the nucleons mean kinetic energy $E_{\mathrm{kin}}\sim\,3T\sim20\MeV$. On the other hand, larger energies cannot be efficiently absorbed by nucleons since their phase space becomes smaller. Therefore, the ALP MFP increases for energies larger than $\sim 20$~MeV. 
 \newline
On the contrary, the MFP associated to pionic processes decreases monotonically as the ALP energy increases. Since these processes consist in the conversion of an ALP into a pion (nucleons can be considered at rest), all the energy brought by the ALP is available in the center of mass to produce the outgoing pion. Consequently, once exceeded the threshold energy $E_a=m_\pi$, as the ALP energy increases the conversion process becomes more and more probable .

\end{document}